\documentclass[epsfig,12pt]{JHEP} 
\usepackage{epsfig} 
 
\newcommand{\al}{\alpha}

\newcommand{\be}{\begin{equation}} 
\newcommand{\ee}{\end{equation}} 
\newcommand{\ba}{\begin{eqnarray}} 
\newcommand{\ea}{\end{eqnarray}} 
\newcommand{\baa}{\begin{eqnarray*}} 
\newcommand{\btab}{\begin{tabular}} 
\newcommand{\etab}{\end{tabular}} 
\newcommand{\eaa}{\end{eqnarray*}} 
 
\newcommand\re[1]{(\ref{#1})} 
 
\def \qqqquad {\qquad\qquad} 
\newcommand\lr[1]{{\left({#1}\right)}}

\newcommand \vev [1] {\langle{#1}\rangle}

\def \e {\mbox{e}} 
\def \CO {{\cal O}}

\def \as {\relax\ifmmode\alpha_s\else{$\alpha_s${ }}\fi} 
\def \MS {\overline{MS}} 
 
\def \GeV {\mbox{GeV}} 
\def \erf {\mbox{erf}} 
 
\title{Nonperturbative corrections to the Drell-Yan transverse momentum distribution  } 
 
\preprint{LPT--Orsay--01--12\\hep-ph/0102237} 
 
\author{  Sofiane ~Tafat 
\\ \vspace{.3in} Laboratoire de Physique 
Th\'eorique\footnote{Unite Mixte de Recherche du CNRS (UMR 8627)}, 
Universit\'e de Paris XI, \\ 
91405 Orsay Cedex, France\\ 
E-mail: \email{Sofiane.Tafat@th.u-psud.fr }} 
 
\abstract{We study nonperturbative corrections to the transverse momentum 
distribution of vector bosons in the Drell-Yan process. 
Factorizing out the Sudakov effects due to soft gluons we express 
their contribution to the distribution in the form of the vacuum 
averaged Wilson loop operator. We calculate the nonperturbative 
contribution to the Sudakov form factor using the expansion of the 
Wilson loop over vacuum fields supplemented with the expression 
for nonlocal gauge invariant field strength correlator. Although 
the Wilson loop is defined in an essentially Minkowski kinematics, 
the part of the nonperturbative contribution depending on the 
invariant mass of the produced vector bosons is governed by 
asymptotics of the correlator at large space-like (Euclidean) 
separations and therefore can be calculated using conventional 
nonperturbative methods. Applying the results of lattice 
calculations we found that the obtained expression for the 
nonperturbative power corrections is in qualitative agreement with 
known phenomenological expressions at large transverse momenta and 
deviate from them at small transverse momenta.}

 0 
 
\begin{document} 
\maketitle 
 
\section{Introduction} 
 
Transverse momentum distributions of the vector bosons produced in 
the Drell-Yan process is one of the classical examples of 
successful QCD description of the phenomenology of hadron-hadron 
processes at high energies \cite{Old,Gaus,CSS}. Increasing 
accuracy of the experimental data combined with the state-of-art 
resummed perturbative calculations allow to test perturbative QCD 
predictions and, at the same time, estimate nonperturbative 
corrections to them associated with the confinement effects. It 
was proposed a long ago to parameterize the latter corrections by 
introducing additional phenomenological parameters into 
perturbative formulas \cite{Gaus,CSS}. Their values were extracted 
from the comparison with the experimental data \cite{LY}. Despite 
the fact that the obtained QCD predictions agree well with the 
experimental data \cite{recent} throughout a wide interval of 
invariant mass of the vector bosons, $Q^2$, and their transverse 
momentum, $q^2$, the understanding of nonperturbative effects in 
the Drell-Yan process is still missing. This should be compared 
with the situation with hadronization corrections to the deeply 
inelastic scattering (DIS) and inclusive observables in 
$e^+e^--$annihilation. There, the operator product expansion (OPE) 
allows to organize the nonperturbative corrections  in inverse 
powers of hard scale $\Lambda_{2n}/Q^{2n}$ and identify the 
corresponding nonperturbative scales $\Lambda_{2n}$ as universal 
matrix element of composite local operators in QCD like gluon 
condensate in $e^+e^--$annihilation and higher twist operators in 
DIS. Since the OPE is not applicable to the Drell-Yan production, 
the physical interpretation of the measured nonperturbative scales 
remains unclear. Different approaches have been proposed to 
describe these scales within QCD \cite{KS,nonPT}. 
 
In this paper we shall follow the approach developed in \cite{KS} 
and consider nonperturbative corrections the transverse momentum 
distribution $d^2\sigma/d Q^2 d q^2$ in the end-point region 
describing the production of the vector bosons with large 
invariant mass $Q^2$ and small transverse momentum $q^2\ll Q^2$. 
It is well known that both perturbative and nonperturbative 
corrections are enhanced in this region. Indeed, perturbative QCD 
analysis shows that the cross section gets large perturbative 
corrections $\alpha_s^n/q^2 \log^{2n-1}(Q^2/q^2)$. These 
corrections are associated with multiple soft and collinear gluon 
emissions and they need to be resummed to all orders in the 
coupling constant \cite{Old}. Applying the Collins-Soper-Sterman 
resummation formalism one gets \cite{CSS} 
\be 
Q^2\frac{d^2\sigma}{dQ^2dydq^2}=\sigma_0\sum_j e_j^2 
\int \frac{d^2b}{(2 \pi)^2}\,\e^{-i\vec b\cdot\vec q}\e^{-S(b^2,Q^2)} 
C_{jh}(x_1,Q,b)C_{jh'}(x_2,Q,b)+ Y(Q,q,x_1,x_2)\,. 
\label{1} 
\ee 
Here, $x_1=Q/\sqrt{s}\e^{y}$ and $x_2=Q/\sqrt{s}\e^{-y}$ are momentum fractions of parton 
with relative rapidity $y$ and the invariant energy of colliding hadrons $s$, 
$\sigma_0=\frac{4\pi\alpha^2}{9s}$ is the Born level partonic cross-section and 
$C_{jh}$ are given by convolution of partonic distributions $f_{a/h}(x)$ with 
perturbatively calculable coefficient functions $c_{ja}$ 
\be 
C_{jh}(x,Q,b)=\sum_a\int_x^1\frac{d\xi}{\xi} c_{ja}(x/\xi,b,Q)f_{a/h}(\xi,b_0/b) 
\ee 
with $b_0=2\e^{-\gamma_E}$ and $\gamma_E$ being the Euler constant. 
Integration over two-dimensional impact parameter $\vec{b}$ ensures that the total 
transverse momentum of the vector boson equals $\vec{q}$. In Eq.~\re{1}, the large 
perturbative contributions due to soft gluons are factored out into the Sudakov form 
factor given by 
\be 
S(b^2,Q^2)=\int_{b_0^2/b^2}^{Q^2} \frac{dk^2}{k^2} 
\lr{A(\as(k^2))\log\frac{Q^2}{k^2}+B(\as(k^2))} 
\label{2} 
\ee 
with the coefficient functions $A(\alpha_s)$ and $B(\alpha_s)$ known to two-loop order 
\cite{Gaus}. The function $Y(Q,q,x_1,x_2)$ describes the  part 
of the cross-section \re{1} regular at $q^2\ll Q^2$. It is 
obtained by subtracting the logarithmically enhanced part from the fixed order result. 
 
The Sudakov form factor \re{2} resummes large perturbative 
logarithms $\alpha_s^n \ln^{2n-m}(Q^2b^2)$ and it suppresses the 
large $b^2$, or equivalently small $q^2$ region of the 
distribution. Although it is defined for arbitrary 
$b^2$ it was realized that the perturbatively resummed formula 
\re{1} should be modified at large $b^2$. 
 
Firstly, the expression \re{2} was found by summing of all large 
logarithms $\log(b^2Q^2)$ in the region $Q^2 \gg 1/b^2 \gg 
\Lambda_{\rm QCD}^2$ where perturbative QCD works. However, since 
in \re{1} the integration goes over all values of $b^2$ the 
modification of the resummation formula is needed for $1/b^2 \sim 
\Lambda^2_{\rm QCD}$. To this end it was proposed \cite{CSS,Gaus} to 
introduce the scale $Q_0$ below which the perturbation QCD 
fails to replace the impact parameter $b$ by the function 
$$ 
b^*=\frac{b}{[1+b^2Q_0^2]^{1/2}} 
$$ 
in the expression for the Sudakov form factor, $S$, and for the 
distribution function, $C$, in \re{1}. Then, at large $b$ thus 
defined function $b^*$ approaches its maximum value $b^* = 1/Q_0$ 
and perturbative expressions remain well defined. 
 
Secondly, we expect nonperturbative corrections become equally 
important at small transverse momentum $q^2$, or large $b^2$. The 
latter corrections appear suppressed by powers of both scales, 
$Q^2$ and $q^2$. However, for small transverse momentum $q^2\ll 
Q^2$ it becomes reasonable to neglect all power corrections on the 
larger scale, $Q^2$, and restrict consideration to nonperturbative 
effects on the smaller scale $q^2$. The standard way to incorporate 
leading nonperturbative effects into \re{1} amounts to replacing 
the Sudakov form factor in \re{1} by the following expression \cite{CSS,Gaus}. 
\be 
S_{\rm PT}(b^2,Q^2) \to S_{\rm PT}((b^*)^2,Q^2) + S_{\rm nonPT}(b^2,Q^2) 
\label{non-S} 
\ee 
with nonperturbative part given by 
\be 
S_{\rm nonPT}(b^2,Q^2)= \phi(b) \ln\frac{Q}{2Q_0} + \phi_{j/h}(b,x_1) + 
\phi_{j/h'}(b,x_2)\,. \label{mod} 
\ee 
Here, $\phi$ and $\phi_{j/h}$ are some phenomenological functions whose form 
should be fitted to the experimental data. The simplest parameterization 
of the functions looks as follows 
\be 
\phi(b)=g_2 b^2, 
\qquad 
\phi_{j/h}(b,x_1)+ \phi_{j/h'}(b,x_2)=g_1 b^2, 
\label{4} 
\ee 
The values of the scales $g_1$ and $g_2$ cannot be 
calculated in perturbative QCD and one finds them from comparison 
with the experimental data as \cite{Gaus,LY} 
\be 
g_1=0.24~^{+0.08}_{-0.07}\ (\GeV)^2, 
\qqqquad 
g_2=0.34~^{+0.07}_{-0.08}\ (\GeV)^2. 
\label{5} 
\ee 
There exist other proposals for the parameterization of nonperturbative effects in 
this process \cite{LY}. 
 
In the present paper we study nonperturbative corrections to the resummation formula 
\re{1} and calculate the power suppressing factor \re{4} using the notion 
of QCD vacuum condensates. 
%
%
The paper is organized as follows. In Sect.~2 we consider the origin of the Sudakov 
form factor in the resummation formula \re{1}. Using factorization properties of soft 
gluons we express this form factor as the expectation value of the Wilson loop operator 
built from soft gluon fields. In Sect.~3 we show that the lowest order perturbative 
calculation of the Wilson loop expectation value reproduces known expression for the 
form factor \re{2}. At the same time, as shown in Sect.~4, to higher orders in 
perturbation theory this expectation value contains ambiguity associated with the 
contributions of 
infrared renormalons having the form of the power correction ${\cal O}(b^2\Lambda^2)$ 
\cite{KS}. In Sect.~5 we calculate the nonperturbative power corrections to the 
Wilson loop by taking into account interaction of partons with vacuum gluon fields 
parameterized by nonlocal QCD condensates. In Sect.~6 we analyze the resummation 
formula for transverse momentum distribution with nonperturbative effects 
taking into account and compare it with the phenomenological ansatz \re{mod}-\re{5}. 
Sect.~7 contains concluding remarks. 
 
\section{Factorization of soft gluons} 
 
The factorization formula \re{1} is valid up to corrections suppressed by a power 
of high scale $Q^2$ and it takes into account both perturbative and nonperturbative 
corrections on a smaller scale $q^2$ for $q^2 \ll Q^2$. The different factors 
entering the factorization formula \re{1} take into account contribution from 
different physical subprocesses - hard, collinear and soft - and have the following 
interpretation. 
All emissions of quarks and gluons with momenta collinear to incoming hadron momenta 
$h$ and $h'$ are factorized into two distribution functions $f_{a/h}(x_1,b_0/b)$ and 
$f_{b/h'}(x_2,b_0/b)$. Hard emissions are described by the coefficient functions 
$c_{jh}$ which, in general, are not universal and depend on the produced vector 
boson. The soft subprocess describes both the initial state interaction between 
partons and final state radiation of soft gluons. It is factorized into the Sudakov 
form factor and will be the central object of our consideration. Its perturbative 
expression is given by \re{2} and we anticipate that nonperturbative effects 
will modify it at the level of power corrections ${\cal O}(\Lambda_{\rm QCD}^{2n} b^{2n})$ 
which we shall sum over all $n$. 
 
Let us demonstrate that the Sudakov form factor can be calculated in QCD as an 
expectation value of the Wilson loop operator built from soft gluon fields \cite{12}. 
Interaction of soft gluons with quarks can be treated using the eikonal approximation. 
In this way, one can obtain the perturbative expression for the Sudakov form factor 
by calculating the Feynman diagrams order by order in perturbation theory and replacing 
quark propagators and quark-gluon vertices by their approximate eikonalized expressions. 
However, in order to go beyond the perturbation theory one should find an operator 
definition of the Sudakov form factor that does not refer to the Feynman diagrams and 
that is given entirely in terms of gluon and quark fields. Such definition exists 
and it is based on the Wilson line operators defined as 
\be 
W[C] = P\exp\lr{ig\int_C dz_\mu A^\mu(z)} 
\label{W} 
\ee 
with $A_\mu(z)$ being the gauge field describing soft gluons and $C$ being an 
arbitrary integration path $C$ in the Minkowski space. 
 
The appearance of the Wilson line operators can be understood as follows \cite{11}. 
It is well known that in the eikonal approximation the interaction of quark with 
soft gluons leads to the appearance of the eikonal phase of the quark wave function 
\be 
\Psi(x) \to \Phi_p[-\infty,x] \Psi(x) 
\label{fact} 
\ee 
with $\Phi_p^\dagger\Phi_p=1$. The phase $\Phi_p[-\infty,x]$ is given by the Wilson 
line operator \re{W} evaluated along the classical trajectory of quark. For quark 
with  momentum $p_\mu$ this trajectory is a straight line 
that can be parameterize as $z_\mu=p_\mu s + x_\mu$ with $s$ being a proper time 
\be 
\Phi_p[-\infty,x] = P\exp\lr{ig\int_{-\infty}^0 ds p_\mu A^\mu(ps+x)}\,. 
\label{phase} 
\ee 
Going over to the momentum representation, it is straightforward to verify that 
\re{fact} and \re{phase} reproduce well known expression for the eikonalized amplitude of 
soft gluon emissions. 
 
In the case of the Drell-Yan process, the Sudakov form factor appears as a 
contribution of real and virtual soft gluons interacting with incoming quark and 
antiquark before their annihilation. Applying the eikonal approximation \re{fact} 
we find that both quark and antiquark acquire eikonal phases which are factorized from 
the amplitude of the process into 
\be 
U_{\rm DY}(x) = T \left\{\Phi_{-p_2}[-\infty,x]\Phi_{p_1}[x,-\infty]\right\}\,, 
\label{U} 
\ee 
where $p_1^\mu$ and $p_2^\mu$ are momenta of quark and antiquark, respectively, 
$x_\mu$ denotes the annihilation point and $T$ stands for time ordering of 
gluon field operators. Projecting out the total eikonal phase \re{U} onto 
all possible final states consisting of an arbitrary number of soft gluons, $N\ge 0$, 
with the total transverse momenta $(-q)$ we can write the soft gluon contribution 
to the Drell-Yan cross-section as 
\be 
\sigma_{\rm eik} = \sum_N |\vev{0|U_{\rm DY}(0)|N}|^2 \delta^{(2)}(\vec k_N + \vec q) 
\nonumber=\int\frac{d^2b}{(2\pi)^2}\ \e^{-i\vec b\vec q}\, W_{\rm DY}(b), 
\label{eik} 
\ee 
where 
\be 
W_{\rm DY}(b)=\sum_N 
\vev{0|U_{\rm DY}(0)|N}\e^{-i\vec b\vec k_N} 
\vev{N|U_{\rm DY}^\dagger(0)|0} 
\ee 
and $|N\rangle$ is the final state of $N$ soft gluon with the total momentum 
$k_N^\mu$. Using the translation invariance property 
$ 
U_{\rm DY}(b) = \e^{iPb} U_{\rm DY}(0) \e^{-iPb} 
$ 
with $P^\mu|N\rangle = k_N^\mu |N\rangle$ being the total momentum operator, one 
can perform summation over all possible intermediate states to get \cite{KS} 
\be 
W_{\rm DY}(b) = \vev{0|U_{\rm DY}(0) U_{\rm DY}^\dagger(b)|0} 
\equiv\vev{0|P\exp\lr{ig\int_{C_{DY}} dz_\mu A^\mu(z)} |0}\,. 
\label{6} 
\ee 
The integration contour $C_{DY}$ entering this expression is shown in Fig.~1. It 
consists of two similar parts having the form of an (infinite) angle which are 
separated in the transverse direction by the impact vector $\vec b$. They belong to 
the plane defined by the quark momenta, $p_1^\mu$ and $p_2^\mu$, and correspond to 
the eikonalized amplitude of the process and its conjugated counterpart. Two cusps 
located at the space-time points $0$ and $b$ define the annihilation points. The 
cusp angle $\chi$ is equal to the angle between quark momenta $p_1$ and $p_2$ in 
Minkowski space 
\be 
\cosh\chi = {(p_1p_2)}/p^2=\frac{Q^2}{2p^2}-1\,, 
\label{chi} 
\ee 
where $Q^2=(p_1+p_2)^2$ is the invariant mass of produced vector boson and 
we put $p_1^2=p_2^2=p^2$ to regularize light-cone singularity that appear 
for $p^2\to 0$. We find that for $Q^2/p^2 \gg 1$ the cusp angle scales as 
$\chi=\ln\frac{Q^2}{p^2}$ and therefore the (kinematical) dependence of the 
Wilson loop on the cusp angle is translated into $\ln Q^2-$dependence of the 
eikonalized cross-section \re{eik}.

\begin{figure}[ht] 
\centerline{\epsfxsize 10cm\epsffile{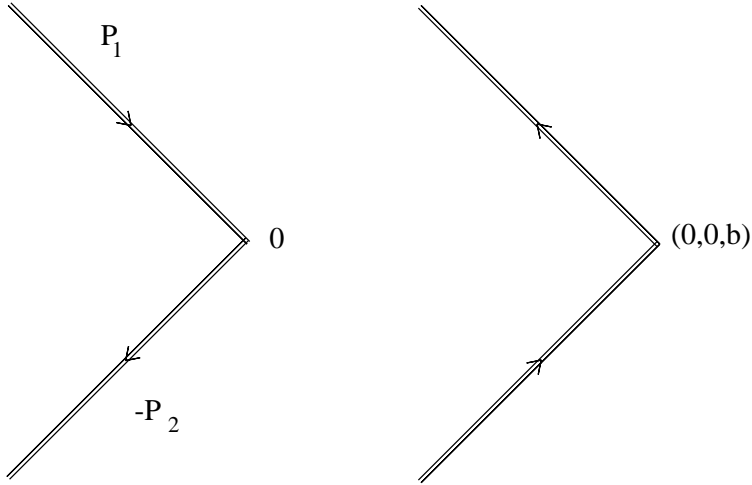}} 
\caption[]{The integration path $C_{\rm DY}$ defining the eikonal phase \re{6}. 
Straightforward lines correspond to classical trajectories of quark and antiquark 
with momenta $p_1$ and $p_2$, respectively.} 
\label{Fig-1} 
\end{figure} 
 
We would like to notice that the Wilson loop in \re{6} should be averaged with 
respect to the state describing the spectator partons of both incoming hadrons. 
In this way one takes into account the possibility for quark and antiquark 
participating at the partonic subprocess to exchange by soft gluons with 
constituents of both hadrons before their annihilation. In what follows, 
we shall neglect such effects for the sake of simplicity 
and average \re{6} with respect to the vacuum state. There are few reasons why such 
an approximation is reasonable. First, it is known \cite{11} that all perturbative effects of 
interaction between hadron constituents and particles participating in short 
distance partonic subprocess cancel in sum of all Feynman diagram up to corrections 
suppressed by a power of the hard scale $Q^2$. Second, analysis of nonperturbative 
corrections to the transverse momentum distributions of jets produced in 
$e^+e^--$annihilation and deep inelastic scattering indicates \cite{MOS} that 
the scale $g_2$ parameterizing nonperturbative corrections in Eqs.~\re{4} and 
\re{5} is approximately universal and does not depend on the initial hadronic state. 
 
\section{Perturbative contribution} 
 
Using the expression \re{6} for the Sudakov form factor we may try to take into account 
both perturbative and nonperturbative effects. The Sudakov form factor depends on $Q^2$, lepton pair 
invariant mass of produced vector boson, and $b^2$, impact parameter conjugated to 
transverse momenta $q$. In the Wilson loop representation \re{6}, this 
dependence follows from the properties of the integration path $C_{DY}$ in Fig.~1. 
In particular, the impact parameter $b$ controls the transverse size of the 
integration contour. As a consequence, putting $\vec b=0$ and using unitarity 
of the eikonal phase one finds that 
\be 
W_{\rm DY}(\vec b=0)=1 
\ee 
provided that all perturbative (light-cone and short-distance) singularities are 
appropriately regularized. This suggests to look for an expansion of the Wilson 
loop expectation value around $\vec b=0$. We expect that perturbative expansion 
will generate $\ln b^2-$corrections whereas nonperturbative effects will give 
rise to corrections in powers of $b^2$. 
 
Let us start with perturbative calculation of \re{6} and show that the expression \re{6} 
reproduces the Sudakov form factor (2) in perturbative QCD at small $b^2$. Using the 
definition \re{6} we get the following expression for the gauge invariant Wilson loop 
to the lowest order of perturbation theory: 
\be 
W^{(1)} = 1+ \frac{(ig)^2}2\int_{C_{DY}}  dx^\mu \int_{C_{DY}} dy^\nu \vev{0|A_\mu(x) A_\nu(y)|0} = 1 
-\frac{g^2}2C_F\int_{C_{DY}} dx^\mu \int_{C_{DY}} dy^\nu D^{\mu\nu}(x-y) 
\label{8} 
\ee 
with $C_F=(N_c^2-1)/(2N_c)$ and $D_{\mu\nu}(x)$ being a free gluon propagator. To 
evaluate this integral we have to fix a gauge for a gluon field and, in addition, 
take care of possible divergencies of the integral. To this end we choose the Feynman 
gauge and introduce the dimensional regularization with $D = 4 -2\varepsilon$ being the 
dimension of the $x-$space. Substituting the gluon propagator in the 
Feynman gauge and integrating it twice over the path $C_{\rm DY}$ we get the following 
expression for $W^{(1)}$ 
\be 
W^{(1)}=1+g^2\mu^{4-D}C_F\int\frac{d^Dk}{(2\pi)^D}2\pi\delta_+(k^2) 
\lr{\frac{p_1}{(p_1k)}-\frac{p_2}{(p_2k)}}^2 
(\e^{i\vec k\cdot\vec b}-1), 
\label{9} 
\ee 
where $\mu$ is a dimensionful parameter of the dimensional regularization, $d^Dk\equiv 
dk_+dk_-d^{D-2}\vec{k}$ with $k_\pm$ being light-cone components and 
$\vec{k}$ being $(D-2)-$dimensional transverse momentum of gluon. Integration over 
$k_\pm$ leads to: 
\be 
W^{(1)} = 1+ 4\as C_F (\chi\coth\chi-1) \mu^{4-D} 
\int\frac{d^{D-2}\vec{k}}{(2\pi)^{D-2}}\frac{\e^{i\vec{k}\cdot\vec{b}}-1}{\vec{k}^2}, 
\label{10} 
\ee 
where the cusp angle $\chi$ was defined in \re{chi}. 
We notice that this integral does not have infrared divergences at small $\vec{k}^2$ 
but it has ultraviolet divergence coming from large transverse momentum. Dimensional 
regularization provides a natural cut-off $\mu$ on maximal momenta of soft gluons and, 
as a consequence, the final expression for $W^{(1)}$ contains a pole in $\varepsilon$ 
\be 
W^{(1)} = 1- \frac{\as}{\pi} C_F (\chi\coth\chi-1) 
\frac{\Gamma(1-\varepsilon)}{\varepsilon} (\pi\mu^2\vec{b}^2)^\varepsilon. 
\ee 
Note that this divergence has nothing to do with conventional UV divergences in 
perturbative QCD since the residue of a pole in \re{10} depends explicitly on the cusp 
angle $\chi$ which, by the definition \re{chi}, is a function of the kinematical 
invariants of the process. This divergence is 
introduced by the factorization procedure which separates the contribution of soft 
gluons from the exact amplitude of the process  \cite{13}. Applying the eikonal approximation 
to the amplitude we correctly describe its IR behavior but change its UV 
asymptotics. Since by the definition only soft gluons contribute to the Wilson loop 
expectation value \re{6}, we should put an upper bound on the value of the transverse 
momentum in \re{10}. In the dimensional regularization this cut-off is given by 
$k_{\rm max}^2=\mu^2$. 
 
Specific UV divergences which we observed to one-loop order are general feature of 
Wilson lines. These divergences, the so-called cusp singularities, appear due 
to the fact that the integration path $C_{\rm DY}$ entering the definition \re{6} of 
the Wilson lines contains cusps with angle $\chi$. The renormalization properties of the cusp 
divergences were studied in detail. It was shown that cusp singularities are 
renormalized multiplicatively and the corresponding anomalous dimension $\Gamma_{\rm 
cusp}(\as,\chi)$ was found to the lowest orders in perturbative QCD 
\cite{13}. 
 
Subtracting UV pole in the $\MS-$scheme we get the one-loop renormalized Wilson line as 
\be 
W^{(1)} = 1- \frac{\as}{\pi} C_F (\chi\coth\chi-1) 
\log\frac{\mu^2\vec{b}^2}{{b}_0^2} 
\label{col} 
\ee 
where the constant ${b}_0$ was defined in \re{2}. Substituting expression \re{chi} 
for the cusp angle $\chi$ we encounter another singularity. Performing calculations of 
the partonic cross sections in perturbative QCD we neglect quark virtualities 
by putting $p_1^2=p_2^2=0$. In this limit the integration path in Fig.~1 goes along the 
light-cone and one expects additional singularities to appear. Indeed, using the 
definition \re{chi} we find that the angle $\chi$ is infinite at $p^2=0$ and the 
one-loop expression for $W^{(1)}$ is divergent. As was shown in \cite{13}, this 
singularity originates from soft gluons propagating along the light-cone. The 
renormalization properties of light-like Wilson lines were studied in \cite{14}. 
It was found that the renormalized Wilson loop having both light-cone and cusp 
singularities obeys the following renormalization group equation: 
\be 
\lr{\mu\frac{\partial}{\partial\mu}+\beta(g)\frac{\partial}{\partial g}} 
\frac{\partial \ln W(b^2\mu^2,Q^2/\mu^2)}{\partial \ln Q^2} = -2 \Gamma_{\rm cusp}(\as) 
\ee 
where $\Gamma_{\rm cusp}(\as)$ is the cusp anomalous dimension 
\be 
\Gamma_{\rm cusp}(\as)=\frac{\alpha_s}{\pi}C_F+{\cal O}(\alpha_s^2). 
\label{cusp} 
\ee 
The general solution to 
this RG equation is given by \cite{12} 
\be 
W(b^2\mu^2,Q^2/\mu^2) = \exp\lr{-\int_{b_0^2/b^2}^{\mu^2}\frac{dk^2}{k^2} 
\lr{\Gamma_{\rm cusp}(\as(k^2))\log\frac{Q^2}{k^2}+\Gamma(\as(k^2))}}W_0(b^2,Q^2) 
\label{fin} 
\ee 
where $\Gamma$ is the integration constant and $W_0(b^2,Q^2)$ is 
the boundary value of the Wilson loop at the normalization scale 
$\mu^2=b_0^2/b^2$. Since $\mu$ has the meaning of the upper limit on the energy 
of soft gluons we can put $\mu^2=Q^2$ in \re{fin}. Then, the resulting expression 
for the Wilson loop expectation value \re{fin} is given by the product of two 
factors. One of them resummes all large perturbative logarithms in $b^2$ coming 
from integration over soft gluon transverse momenta above the IR scale $1/b^2$. 
It coincides with the Sudakov form factor \re{2} provided that the coefficient 
functions $A(\as)$ and $B(\as)$ are given by the anomalous dimensions, $A(\as) 
=\Gamma_{\rm cusp}(\as)$ and $B(\as) =\Gamma(\as)$. 
 
\section{IR renormalons in Wilson loops} 
 
The solution to the evolution equation \re{fin} is defined up to the factor 
$W_0(b^2,Q^2)$ which gets contribution from soft gluons with transverse momentum 
below the scale $1/b^2$ and, therefore, cannot be calculated in perturbative QCD. 
It is this factor that provides nonperturbative contribution to the transverse 
momentum distribution, Eqs.~\re{non-S} and \re{5}, and that we expect to be 
of the following form 
\be 
W_0(b^2,Q^2) = \exp\lr{\phi(b)\ln\frac{Q}{2Q_0}+ {\cal O}(\ln^0 Q)}\,. 
\label{W0-IR} 
\ee 
Here, $\phi(b)$ is some function of the impact parameter $b^2$ which can not be 
calculated perturbatively. In order to justify \re{W0-IR} and to determine the 
explicit form of the function $\phi(b)$ one has to perform a nonperturbative 
calculation of the Wilson loop expectation value \re{6}. This will be done in the 
next Section. However, before going through an explicit calculations it becomes 
instructive to examine the general structure of nonperturbative corrections 
to $W_0(b^2,Q^2)$  by calculating the contribution of the IR renormalons to the 
Wilson loop expectation value \re{6}. 
The IR renormalons describe perturbative contribution to $W(b^2\mu^2,Q^2/\mu^2)$ 
coming from soft gluon with momenta $k\sim\Lambda_{\rm QCD}$. The later turns out to 
be ambiguous at the level of power corrections $\sim \exp(-n\beta_0/\alpha_s(1/b^2)) = 
(b^2\Lambda_{\rm QCD}^2)^n$ with the exponent $n$ uniquely fixed by 
large order behaviour of the perturbative series. For the physical 
result \re{fin} for the Wilson loop to be uniquely defined, this ambiguity should be absorbed 
into the nonperturbative power correction to the boundary condition $W_0(b^2,Q^2)$. 
 
To estimate the IR renormalon contribution to the Wilson loop we shall apply the 
renormalon technique and calculate \re{6} in the so-called single renormalon 
chain ($1-$chain) approximation \cite{Beneke}. Applying the Borel transformation we can 
write the perturbative contribution to the Wilson loop as 
\be 
W^{(\rm 1-chain)}-1=\int_0^\infty du \exp(-u/\alpha_s\beta_0) B[W](u) 
\label{1-ch} 
\ee 
where the Born term was subtracted in the l.h.s. Calculating 
$W^{(\rm 1-chain)}$ we start with the lowest-order expression for the 
Wilson loop, replace the gluon propagator by the chain of 
fermionic bubbles and then apply naive nonabelianization to promote 
the quark contribution to the full one-loop expression for the 
$\beta-$function. In this way, one finds the following 
representation for the Borel transform \cite{Beneke} 
\be 
B[W](u) = -\frac{\sin(\pi u)}{\beta_0\pi u \al_s}\int_0^\infty d\lambda^2 
\lr{\frac{\lambda^2\e^{C}}{\mu^2}}^{-u}\frac{d}{d\lambda^2} 
\, W^{(\rm 1-chain)}(\lambda^2) 
\label{BW} 
\ee 
where $C$ is a renormalization scheme constant and $C=-5/3$ in 
the $\overline{\rm MS}-$scheme. Here, $\lambda^2$ is the dispersive 
parameter which enters into perturbative expression for the Wilson 
loop in the form of a fictitious gluon mass. This allows us to 
write $W^{(\rm 1-chain)}(\lambda^2)$ in the form \re{9} with 
$\delta_+(k^2)$ replaced by $\delta_+(k^2-\lambda^2)$. Integration 
over longitudinal components $k_\pm$ leads to the expression 
\be 
W^{(\rm 1-chain)}(\lambda^2) = 1+ 4\as C_F (\chi\coth\chi-1) \mu^{4-D} 
\int\frac{d^{D-2}\vec{k}}{(2\pi)^{D-2}}\frac{\e^{i\vec{k}\cdot\vec{b}}-1} 
{\vec{k}^2+\lambda^2}, 
\label{bad} 
\ee 
which should be compared with \re{10}. Similar to \re{col}, it suffers from collinear 
singularities as $p^2\to 0$, or $\chi\to\infty$. To avoid these 
singularities it suffices to consider the following logarithmic 
derivative 
\be 
\frac{d}{d\ln Q^2}W^{(\rm 1-chain)}(\lambda^2) = C_F 
\frac{\alpha_s}{\pi} \lr{ 2\lr{ \frac{2\pi b \mu^2 }{\lambda}}^\varepsilon 
K_\varepsilon(b\lambda) - \lr{\frac{4\pi \mu^2}{\lambda^2}}^\varepsilon \Gamma(\varepsilon )} 
\label{K} 
\ee 
where $K$ is the Bessel function of the second kind and $D=4-2\varepsilon$. Then, 
substituting \re{K} into \re{BW} we obtain after some algebra 
\be 
B\left[\frac{d W}{d \ln Q^2}\right](u) = -\frac{C_F}{\beta_0\pi} 
(\pi b^2\mu^2)^\varepsilon \lr{b^2\mu^2/4 \e^C}^u 
\frac{\Gamma(1-u-\varepsilon)}{(u+\varepsilon)\Gamma(1+u)}. 
\label{Borel} 
\ee 
This expression for the Borel transform of the Wilson loop in the 
single chain approximation coincides with the similar expression obtained in \cite{KS} 
using different approach. Finally, substituting \re{Borel} into \re{1-ch} and 
assuming that the contribution of a single renormalon chain 
exponentiates one arrives at 
\be 
\frac{d\ln W}{d\ln Q^2}=-\frac{C_F}{\pi\beta_0}(\pi\mu^2\vec b^2)^\varepsilon 
\int_0^\infty 
\frac{du}{u+\varepsilon} 
\frac{\Gamma(1-u-\varepsilon)}{\Gamma(1+u)} 
\lr{{\Lambda^2_{\rm QCD}\vec b^2}/4 \e^C}^{u} 
\label{Borel-1} 
\ee 
 
Let us analyse the Borel  singularities of \re{Borel} and \re{Borel-1}. We 
notice that at $\varepsilon\to 0$ the r.h.s.\ of \re{Borel} 
contains a pole at $u=0$. It has a UV origin and comes from 
integration over large transverse momentum in \re{bad}. As was explained in 
the previous section, the appearance of this pole is an artefact 
of the factorization procedure used to separate the contribution 
of soft gluon. Away from small $u$ we put $\varepsilon=0$ and observe 
that \re{Borel} and \re{Borel-1} acquires a series of IR renormalon poles located 
at integer positive $u$. This suggests to divide the integration region in 
\re{Borel-1} into two parts, $0\le u \le 1/2$ and $u>1/2$, and look for 
the solutions to \re{Borel-1} in the following form 
\be 
\ln W = \ln W^{(\rm PT)} + \ln W^{(\rm IRR)} 
\label{de} 
\ee 
with $W^{(\rm PT)}$ getting a well-defined contribution from small $u$ 
and $W^{(\rm IRR)}$ taking into account IR renormalon poles. 
 
Integrating over small $u$ in \re{Borel-1} we expand the integrand 
in powers of $u$ to get 
$$ 
\frac{d\ln W^{(\rm PT)}}{d\ln Q^2} = \frac{C_F}{\pi\beta_0} 
\sum_{n=1}^\infty (-1)^n\frac{(\pi\mu^2b^2\e^{\gamma_E})^\varepsilon}{\varepsilon^n} 
\int_0^{1/2} du u^{n-1} (\Lambda_{QCD}^2 b^2/b^2_0 \e^C)^u 
$$ 
Subtracting poles in $\overline{\rm MS}-$scheme and summing up the 
series we find 
\be 
\frac{d\ln W^{(\rm PT)}}{d\ln Q^2}=\frac{C_F}{\pi\beta_0} 
\log\frac{\ln(b_0^2 \e^C/\Lambda_{QCD}^2\vec b^2)}{\ln(\mu^2 \e^C/\Lambda_{QCD}^2)} 
\label{17} 
\ee 
where we assumed that $\mu^2,1/b^2\gg \Lambda^2_{QCD}$. It is easy to 
check that the perturbative contribution to \re{fin} satisfies this evolution 
equation and, as a consequence, $W^{(\rm PT)}$ coincides with the first 
factor in the r.h.s.\ of \re{fin}. 
 
Integrating in \re{Borel-1} for $u>1/2$ we deform the integration 
contour around integer positive $u=n$ and estimate the contribution of the IR 
renormalons to the Wilson loop as 
\be 
\frac{d\ln W^{(\rm IRR)}}{d\ln Q^2}=\sum_{n=1}^\infty 
c_n^{\rm IRR}(\Lambda_{QCD}^2\vec b^2/b_0^2)^n 
\equiv \frac12 \phi^{\rm IRR}(b) 
\label{IR-ren} 
\ee 
with arbitrary coefficients $c_n^{\rm IRR}$ depending on the 
prescription that one uses integrating the Borel poles and the 
thus defined $\phi^{\rm IRR}(b)$ being a function of $b^2$ 
regular at $b=0$. Comparing \re{de} with 
\re{fin} we notice that in order for the Wilson line expectation value $W$ to 
be well-defined, the IR renormalon ambiguities of $W^{(\rm IRR)}$ should be 
absorbed into the definition of the boundary value $W_0(b^2,Q^2)$. 
The general form of the IR renormalon contribution can be obtained from 
\re{IR-ren}. It matches \re{W0-IR} and $\phi^{\rm IRR}(b)$ provides IR renormalon 
contribution to the nonperturbative function $\phi(b)$. 
 
 
\section{Wilson loop in vacuum fields} 
 
As was shown in the previous section, we expect nonperturbative effects to 
contribute to the power $\CO(b^2\Lambda^2)$ corrections to the Sudakov form factor. 
To estimate these corrections we shall use the representation of the Sudakov 
form factor as Wilson loop vacuum expectation value \re{6}. The calculation of Wilson 
loops in QCD has a long history. It is known that for small (Euclidean) size contours 
we may apply the perturbation theory to calculate the Wilson loop, while for large 
contours we expect essentially nonperturbative ``area law'' behavior. There were 
attempts to justify this behavior using different models of the QCD vacuum \cite{15,16}. 
Trying to apply these approaches to the particular Wilson loop shown in Fig.~1 we 
notice the following important differences. 
First, the specific feature of the Wilson loop in Fig.~1 is that it has 
essentially Minkowski light-cone geometry whereas the calculation of a 
Wilson loop is usually performed in the Euclidian QCD. 
Second, the integration contour in Fig.~1 is formally of an infinite size, but, 
at the same times, its definition involves a natural small parameter $~\vec b^2$. 
At $\vec{b} = 0$ the integration path is shrunk into a point and the 
Wilson loop takes trivial value \re{6}. Then, for small $b^2\ll 1/\Lambda_{\rm QCD}^2$ 
the perturbation theory provides a meaningful approximation to $W(b)$ whereas for 
$b^2\sim 1/\Lambda_{\rm QCD}^2$ nonperturbative corrections become important. 
 
Following \cite{15,16}, we shall assume that for small $b^2$ 
nonperturbative corrections to $W(b^2)$ come from interaction with vacuum 
fields. 
To evaluate the Wilson loop \re{6} we have to fix a gauge and then express the 
correlator of vacuum gauge fields in terms of gauge invariant quantities. 
This can be done most easily in the Fock-Schwinger gauge or fixed point gauge 
$$ 
(x-x_0)\cdot A(x) = 0. 
$$ 
The special feature of this gauge is that the gauge potential can be expressed in 
terms of field strength tensor $F^{\mu\nu}$ as 
\be 
A^\mu(x)=-(x-x_0)_\nu\int_0^1 d\alpha\ \alpha F^{\mu\nu}(x_0+\alpha(x-x_0)). 
\label{18} 
\ee 
Here, the fixed point $x_0$ is a free parameter of the gauge and the $x_0-$dependence 
should cancel in gauge invariant quantities. To simplify the calculations we choose 
\be 
x_0^\mu=\xi b^\mu 
\ee 
with $\xi$ being an arbitrary gauge parameter. We will demonstrate that 
the final expression for the Wilson loop does not depend on $\xi$. 
 
Let us introduce the parameterization of the integration path of Fig.~1 as follows: 
$C_{\rm DY} = C_I + C_{II} + C_{III} + C_{IV}$, where 
\be 
C_I = p_1 s\,,\quad C_{II} = - p_2 t\,,\quad C_{III} = b-p_1 t\,, 
\quad C_{IV} = b + p_2 s\,, 
\label{19} 
\ee 
as $-\infty < s \le 0$ and $0 \le t < \infty$. 
Then, using the relation \re{18}-\re{19} we evaluate the integral 
$$ 
ig \int_{C_I} dx \cdot A (x) = ig \int_{-\infty}^0 ds\,p_1^\mu A_\mu(p_1 s) = ig 
\xi\int_{-\infty}^0 ds \int^1_0 d\alpha \alpha p^\mu_1 b^\nu F_{\mu\nu}(p_1 s\alpha 
+ b\xi(1-\alpha )) 
$$ 
and changing the integration variables $s\to -s/\alpha$ and $\alpha\to 1-\alpha/\xi$ we 
obtain the relation 
\be 
ig \int_{C_I} dx \cdot A (x)=ig\int_0^\infty ds \int_0^{\xi} d\alpha p_1^\mu b^\nu 
F_{\mu\nu}(-p_1s+b\alpha) 
\label{C1} 
\ee 
in which $\xi-$dependence enters through the upper integration limit. 
Similar calculation of the integral along the part $C_{II}$ , $C_{III}$ and $C_{IV}$ 
leads to the following expressions 
\ba 
ig \int_{C_{II}} dx \cdot A (x)&=&-ig\int_0^\infty dt \int_0^{\xi} d\alpha p_2^\mu 
b^\nu F_{\mu\nu}(-p_2t+b\alpha) 
\nonumber\\ 
ig \int_{C_{III}} dx \cdot A (x)&=&ig\int_0^\infty dt \int_{\xi}^1 d\alpha p_1^\mu 
b^\nu F_{\mu\nu}(-p_1t+b\alpha) 
\nonumber\\ 
ig \int_{C_{IV}} dx \cdot A (x)&=&-ig\int_0^\infty ds \int_{\xi}^1 d\alpha p_2^\mu 
b^\nu F_{\mu\nu}(-p_2s+b\alpha) 
\label{C2} 
\ea 
which depend separately on $\xi$. Combining together the expressions \re{C1} and 
\re{C2}, we obtain the following relation 
\be 
ig \int_{C} dx \cdot A (x)=ig\int_0^\infty ds \int_0^1 d\alpha\,  b^\nu 
\lr{p_1^\mu F_{\mu\nu}(-p_1s+b\alpha)-p_2^\mu F_{\mu\nu}(-p_2s+b\alpha)} 
\label{circ} 
\ee 
in which the dependence of the gauge parameter $\xi$ disappears. 
 
Finally, substituting \re{circ} into \re{8} we find that to the lowest order in 
a gauge field expansion the vacuum averaged Wilson loop is given by 
\be 
W=1-\frac{1}{4N} 
\left\langle 0 \left|g^2 
\left[ 
\int_0^\infty ds \int_0^1 d\alpha\ b^\nu 
\left( 
p_1^\mu F^a_{\mu\nu}(-p_1s+b\alpha)-p_2^\mu F^a_{\mu\nu}(-p_2s+b\alpha) 
\right) 
\right]^2 
\right| 0 \right\rangle 
\label{21} 
\ee 
Although this expression is $\xi-$independent, it is not gauge invariant. To 
restore the gauge invariance, two field strength tensors should be connected 
by a Wilson line to form a nonlocal correlator 
$F^a_{\mu\rho}(x)[P\exp(ig \int_y^x dz \cdot A^c(z)t^c)]_{ 
ab}F^b_{\nu\lambda}(y)$ with the group generators $t^a$ defined in the adjoint 
representation of the $SU(N)$ group. Here, the integration path can be an arbitrary 
and we choose a straightforward line for simplicity. Any other choice leads to the 
expression for the Wilson loop which is different from \re{21} by terms containing 
higher powers of gauge fields. Thus defined gauge invariant 
field strength correlator depends on the difference $(x-y)$ 
and it can be decomposed onto two Lorentz tensors \cite{15,18}: 
\ba 
\langle 0|g^2 F_{\mu\rho}^a(x) 
[P\e^{ig\int^{x}_{0}dz \cdot A(z)}]_{ab} F_{\nu\lambda}^b(0)|0\rangle 
&=&(g_{\mu\nu}g_{\rho\lambda}-g_{\mu\lambda}g_{\nu\rho}) 
 \Phi_1(-x^2+i\epsilon) 
\label{22} 
\\ 
&+&( g_{\rho\alpha}\partial_\mu - g_{\mu\alpha}\partial_\rho) 
 ( g_{\lambda\alpha}\partial_\nu - g_{\nu\alpha}\partial_\lambda) 
 \Phi_2(-x^2+i\epsilon) 
\nonumber 
\ea 
Here, $\Phi_1$ and $\Phi_2$ are gauge invariant functions that receive both 
perturbative and nonperturbative corrections. In the special case of abelian gauge 
group the Bianchi identity $\epsilon^{\mu\nu\rho\lambda} \partial_\mu F_{\nu\rho} = 0$ 
implies that $\Phi_1(-x^2 )=0$. Although it is not the case for QCD, this property 
leads to vanishing the lowest order perturbative QCD correction to $\Phi_1$. 
It is convenient to introduce the following representation 
\ba 
\Phi_1(-x^2+i\epsilon) 
&=&-i\int_0^\infty d\sigma\e^{-i\sigma x^2}\tilde\Phi_1(i\sigma) , 
\nonumber 
\\ 
\frac{d\ \Phi_2(-x^2+i\epsilon)}{d\ x^2}&=& 
-i\int_0^\infty d\sigma\e^{-i\sigma x^2}\tilde\Phi_2(i\sigma)  \,, 
\label{24} 
\ea 
where we took into account that eq.~\re{22} defines the function $\Phi_2$ up to 
an arbitrary constant and it is the derivative of this function that should enter 
into the final expressions for the Wilson loop. The correlator \re{22} receives 
both perturbative and nonperturbative corrections. To the lowest order of 
perturbation theory we use the definition \re{22} to calculate the functions 
$\tilde\Phi_1(\sigma)$ and $\tilde\Phi_2(\sigma)$ as 
\be 
\tilde\Phi_1^{(\rm PT)}(\sigma)=0\,,\quad \tilde\Phi_2^{(\rm PT)}(\sigma) = - 
\frac{g^2}{4\pi^{D/2}}(N^2-1)\, \mu^{4-D} \sigma^{D/2-1}\,. 
\label{PT-Phi} 
\ee 
 
Let us substitute the correlator \re{22} into the expression \re{21} for the Wilson 
loop. Contracting the Lorentz indices, using the identities $(p_1\cdot 
b) = (p_2\cdot b) = 0$ and performing integration over $s$ in \re{21} one arrives 
at 
\be 
(p_1 p_2 ) \int^1_0 ds\,dt\,\e^{-i\sigma(p_1 s+p_2t)^2} = \frac1{2i\sigma} 
\chi\coth\chi 
\ee 
where $\chi$ is the angle between quark and antiquark momenta defined in \re{chi}. 
Here, we put quark momenta off-shell, $p_1^2=p_2^2\neq 0$, in order to avoid light-cone 
singularities. The remaining integral in \re{21} is given by 
\be 
f(i\sigma b^2 ) \equiv i\sigma b^2 \int^1_0 d\alpha \int^1_0 d\beta\, \e^{-i\sigma b^2 ( 
\alpha - \beta )^2} 
\ee 
and it can be expressed in terms of the error function 
${\rm erf}( x ) = \frac2{\sqrt{\pi}}\int_0^x dt \,\e^{-t^2}$ as 
\be 
f(x) 
=2x\int_0^1d\alpha\,(1-\alpha)\,\e^{-x\alpha^2} 
= \e^{-x} - 1 + \sqrt{\pi x}\, 
\erf ( \sqrt{x} ) 
\label{f-def} 
\ee 
for an arbitrary $x$. Finally, we obtain the following expression for the Wilson loop 
\re{21} 
\be 
W=1-\frac{i}{4N}(\chi\coth\chi-1) 
\int_0^\infty\frac{d\sigma}{\sigma^2} 
\left[ 
\tilde\Phi_1(i\sigma)f(i\sigma\vec b^2) 
-2 \tilde\Phi_2(i\sigma)(1-\e^{-i\sigma\vec b^2})\right], 
\label{25} 
\ee 
where the functions $\tilde\Phi_1$ and $\tilde\Phi_2$ parameterize the correlator 
\re{22}. The following comments are in order. 
 
The dependence of the Wilson loop on the invariant mass $Q^2$ and impact parameter 
$b^2$ is factorized in \re{25} into two different factors. The $Q^2-$dependence 
is given by a ``cusp factor'' $(\chi\coth\chi-1)$ which coincides with an 
analogous factor in the perturbative expression \re{10}. At the same time, to evaluate 
the $b^2-$dependence of $W$ we need to know nonperturbative functions 
$\tilde\Phi_1$ and $\tilde\Phi_2$. Notice that in the original expression for the 
Wilson loop \re{21} the nonlocal gluon correlator was integrated over both space-time 
and like-like separations $x^2$ corresponding to the distance between different 
points on the contour $C_{\rm DY}$, whereas the final expression \re{25} contains the 
integral over parameter $\sigma$ and depends on the space-like vector $b$. 
This suggests to perform the Wick rotation: $\sigma\to -i\sigma$ and rewrite \re{25} 
in the limit $Q^2\to\infty$ as 
\be 
\frac{d\ln W}{d\ln Q^2}=-\frac{1}{4N} 
\int_0^\infty\frac{d\sigma}{\sigma^2} 
\left[ 
\tilde\Phi_1(\sigma)f(\sigma\vec b^2) 
-2 \tilde\Phi_2(\sigma)(1-\e^{-\sigma\vec b^2})\right]. 
\label{new} 
\ee 
The advantage of this representation with respect to \re{25} is that the 
light-cone singularities cancel out in the r.h.s.\ of \re{new} and therefore it 
has a well defined limit as quark momenta go on-shell, $p_1^2=p_2^2=0$. 
 
The above mentioned properties of the Wilson loop \re{25} and \re{new} can be 
understood as follows. As was shown in the previous section, leading 
nonperturbative $\sim b^2$ corrections to the Wilson loop come from the region of 
small transverse momenta whereas collinear logs $\sim\ln Q^2/p^2$ appear from 
integration over small angles between gluon momentum and one of the quark momenta 
$p_1$ and $p_2$. Since integration over small angles and small transverse momenta 
can be performed independently, the $Q^2-$dependence of perturbative and 
nonperturbative expressions, Eqs.~\re{10} and \re{25}, respectively, coincide. 
The remaining integration over transverse momenta of gluon is translated into 
$\sigma-$integration in \re{25} with nonperturbative corrections absorbed into 
the functions $\widetilde\Phi_1(\sigma)$ and $\widetilde\Phi_2(\sigma)$. 
 
Expression \re{8} defines first two terms in the expansion of the Wilson loop 
over vacuum gauge field. Natural question arises about the contribution of the 
remaining terms in this expansion. It can be estimated by applying the nonabelian 
exponentiation theorem \cite{19} which allows to distinguish higher order terms 
according to their color structure. It leads to the following general expression 
for the Wilson loop 
\be 
 W = \exp(w_1 + w_2 + \ldots)
 \,, 
\label{web} 
\ee 
where $w_n$ is the total contribution to the Wilson loop $W$ with ``maximally 
nonabelian'' color factors, $w_n\sim C_F N_c^{n-1}$, the so-called ``webs''. 
Expanding this expression one can express the contribution to the Wilson loop 
with arbitrary color factor in terms of maximally nonabelian webs which can be 
defined without reference to a perturbative expansion. Comparing \re{web} with 
\re{25} we find that obtained expression for $W$ provides a lowest order 
contribution to the web $w_1$. Then, applying the nonabelian exponentiation 
theorem we may exponentiate the lowest order expression \re{25} and write the 
Wilson loop in the form \re{web} modulo corrections coming from higher webs. 
According to \re{10}, $w_1$ has the following asymptotics in large $Q^2$ limit: 
$w_1 = \CO(\ln Q^2)$. It can be shown following \cite{13}, that this is a general 
property of the webs corresponding to the Wilson loop shown in Fig.~1, $w_n\sim 
\ln Q^2$. Therefore, being combined together in the exponent of \re{web}, higher 
order webs $w_n$ renormalize the coefficient in front of collinear logarithm 
$\sim\ln Q^2$ in lowest web contribution $w_1$. This property implies that the 
derivative $d\ln W/d\ln Q^2$ does not contain collinear logs and, therefore, it 
is $Q^2-$independent as $Q^2\to \infty$ in agreement with \re{new}. 
 
Substituting \re{PT-Phi} into \re{new} and subtracting pole in the $\bar{MS}-$scheme, 
one finds that perturbative contribution to the 
Wilson loop satisfies the following evolution equation 
\be 
\frac{d \ln W^{(\rm PT)}(Q^2b^2 /b_0^2)}{d\ln Q^2} = - \int^{Q^2}_{b_0^2/b^2} 
\frac{dk^2_\perp}{k^2_\perp}\Gamma_{\rm cusp}(\as(k^2_\perp)) 
\label{EQ} 
\ee 
where $\Gamma_{\rm cusp}$ given by \re{cusp}. The higher order webs provide a higher 
order perturbative contribution to the anomalous dimension $\Gamma_{\rm cusp}$. 
Integrating \re{EQ} one reproduces perturbative contribution to \re{fin}. Applying 
the nonabelian exponentiation theorem to calculate nonperturbative contribution 
to the Wilson loop we arrive at the evolution equation \re{new} which is valid up 
to higher web contribution. The evolution equation \re{new} allows to calculate 
the coefficient in front of $\ln Q^2$ but it does not fix the corresponding scale 
$Q_0^2$ 
\be 
W_0(b^2,Q^2)=\exp\lr{ -\frac{1}{4N}\log\frac{Q^2}{Q_0^2} 
\int_0^\infty\frac{d\sigma}{\sigma^2} \left[ \tilde\Phi_1(\sigma)f(\sigma\vec 
b^2) -2 \tilde\Phi_2(\sigma)(1-\e^{-\sigma\vec b^2}) \right] }. 
\label{28} 
\ee 
This expression was first obtained in \cite{un}. It is valid up to higher order 
webs contribution and it is assumed that perturbative contribution is subtracted 
from the functions $\Phi_1$ and $\Phi_2$. One finds similar expression for the 
Wilson loop using the stochastic model for the QCD vacuum \cite{15}. There, it 
corresponds to the neglecting the contributions of higher cumulants of gauge 
strength fields. 
 
\section{Asymptotic behavior of Wilson loop} 
 
Let us consider the properties of nonperturbative contribution to the Wilson loop 
\re{28}. As follows from \re{28}, the $b^2-$asymptotics of the exponent is 
determined by the properties of the function $f(\sigma b^2)$ defined in \re{f-def}. 
 
\subsection{Small $b^2$ behavior} 
 
Using the definition \re{24}, we find that $f(x)$ takes positive values for $x > 
0$ and for small $x$ it has the following asymptotics 
\be 
f(x)=x-\frac{x^2}{6}+\frac{x^3}{30}-\frac{x^4}{168}+\ldots ,\qquad 
\mbox{for}\ x \ll 1 \,. 
\label{29} 
\ee 
Thus, the exponent of the Wilson loop \re{28} admits the following expansion in 
powers of $b^2$ 
\be 
W_0 = \exp\lr{-\ln\frac{Q^2}{Q_0^2}\left[ c_1b^2 + c_2b^4 + c_3 b^6 + 
\CO(b^8)\right]}, 
\label{small} 
\ee 
where the coefficients $c_n$ are defined as moments of the functions 
$\tilde\Phi_1(\sigma)$ and $\tilde\Phi_2(\sigma)$: 
\ba 
c _1 &=& \frac1{4N} \int_0^\infty \frac{d\sigma}{\sigma} 
\lr{\tilde\Phi_1(\sigma)-2\tilde\Phi_2(\sigma)} 
\nonumber 
\\ 
c _2 &=& - \frac1{24 N} \int_0^\infty d\sigma 
\lr{\tilde\Phi_1(\sigma)-6\tilde\Phi_2(\sigma)} 
\\ 
c_3 &=& \frac1{120 N} \int_0^\infty d\sigma \sigma 
\lr{\tilde\Phi_1(\sigma)-10\tilde\Phi_2(\sigma)} 
\nonumber 
\ea 
The expansion \re{small} is in agreement with analysis of the IR renormalon 
ambiguities of perturbative series for $W$ performed in Sect.~4. Dimensionful 
parameters $c_n$ are related to the nonlocal correlator \re{22}. The explicit 
form of these relations can be found from the definition \re{24} as 
\ba 
c_1 &=& \frac1{4 N}\int_0^\infty d \vec x^2 \left[ 
\Phi_1(\vec x^2) + 2\Phi_2'(\vec x^2)\right] 
\label{c1} 
\\ 
c_2 &=& - \frac1{24 N}\left[\Phi_1(0) + 6\Phi_2'(0)\right] 
\\ 
c _3 &=& - \frac1{120 N}\left[\Phi_1'(0) + 10\Phi_2''(0)\right], 
\ea 
where prime denotes differentiation with respect to $x^2$. These equations imply 
that all dimensionful coefficients $c_n$ except of $c_1$ are related to the short 
distance asymptotics of the nonlocal correlator \re{22}. Expanding the l.h.s. of 
\re{22} in powers of $x^2$ one can calculate the first few terms of the small 
$x^2$ expansion of the functions $\Phi_1(-x^2)$ and $\Phi_2(-x^2)$ in terms of 
local vacuum condensates as 
\ba 
\Phi_1(-x^2) &=& \frac{\vev{g^2 F^2}}{12} + x^2\frac{\vev{g^3 f F^3}}{48} + 
\CO(x^4), 
\\ 
\frac{d \Phi_2( - x^ 2 )}{d x^2} &=& - x^ 2\frac{\vev{g ^4 j ^2} }{288} + 
\CO(x^4) 
\ea 
where $\vev{g^ 3 f F^ 3} \equiv \vev{g^ 3 f^{abc} F^ a_{\mu\nu} F^b_{\nu\rho} 
F^c_{\rho\mu}}$, $\vev{ g^ 4 j^ 2}  \equiv \vev{ g^ 4 j^a_\mu j^a_\mu}$ and 
$j^a_\mu$ is a quark current. Using these relations we evaluate the coefficients 
$c_2$ and $c_3$ as 
\be 
c_2=-\frac{\vev{g^2F^2}}{288N}, \qquad c_3=\frac{1}{1152N}\lr{\frac{\vev{g^3fF^3}}{5} 
-\frac{\vev{g^4j^2}}{3}} 
\label{34} 
\ee 
We would like to stress that, in general, the scale $c_1$ is not related to local 
vacuum condensates.%
\footnote{This is in agreement with the fact that there are no local gauge 
invariant operators in QCD of the dimension $2$.} To find its value one needs to 
know the behavior of the correlator \re{22} at large (Euclidean) distances. 
 
\subsection{Large $b^2$ behavior} 
 
The expansion \re{29} and \re{small} has a finite radius of convergence and it is not applicable 
at large $x$, or equivalently large impact vectors $\vec b^2$. Using the 
definition \re{24} we find the asymptotic behaviour of the function $f(x)$ at 
large $x$ as 
\be 
f(x)=\sqrt{\pi x}-1+\e^{-x}\lr{\frac{1}{2x}-\frac{3}{4x^2}+\ldots} ,\qquad 
\mbox{for}\ x\gg 1. 
\label{30} 
\ee 
Its substitution into \re{28} yields 
\be 
W_0(b^2,Q^2)=\exp \lr{-\ln\frac{Q^2}{Q_0^2}\left[c_0\sqrt{\vec 
b^2}+\CO(b^0)\right]}, 
\ee 
where the dimension $1$ coefficient $c_0$ is defined as 
\be 
c_0=\frac{\sqrt{\pi}}{4N}\int_0^\infty d\sigma\,\sigma^{-3/2} \tilde\Phi_1(\sigma) 
 =\frac1{2N}\int_0^\infty d\vec x^2 
(\vec x^2)^{1/2} \Phi_1(\vec x^2). 
\label{30-0} 
\ee 
We notice that the function $\Phi_2(x^2)$ does not contribute to $c_0$ and the 
large $b^2$ behavior of the Wilson loop is governed entirely by the function 
$\Phi_1(x^2)$. 
 
Summarizing, we conclude that the Wilson loop has different behavior at small and 
large impact vectors $b^2$: 
\be 
\ln W_0(b^2,Q^2)= -\ln\frac{Q^2}{Q_0^2}\, c_1b^2\,,\qquad \mbox{for 
$b^2\ll\lambda^2$} 
\label{sol-1} 
\ee 
and 
\be 
\ln W_0(b^2,Q^2)= -\ln\frac{Q^2}{Q_0^2}\, c_0\sqrt{b^2}\,,\qquad \mbox{for 
$b^2\gg\lambda^2$} 
\label{sol-2} 
\ee 
where $\lambda$ is the characteristic length defined by the properties of the 
nonlocal correlator \re{22}. The nonperturbative scales $c_0$ and $c_1$ are given 
by \re{30-0} and \re{c1}, respectively.  We recall that the expressions \re{28}, 
\re{sol-1} and \re{sol-2} were obtained by solving the evolution equation \re{new} 
and therefore they are 
valid up to corrections independent on $Q^2$. In the expressions \re{mod} and 
\re{4} the later 
corrections are parameterized by the scale $g_1$. 
 
\subsection{Ansatz for the nonlocal correlator} 
 
Calculating the scales $c_1$ and $c_0$ we shall rely on particular 
nonperturbative ansatz for the nonlocal correlator \re{22}. Since their values 
depend on the behavior of the nonlocal correlator \re{22} at large euclidian 
distances $x^2 < 0$, we may apply the results of lattice calculations of this 
correlator. 
 
According to \cite{G,15}, the nonlocal correlator \re{22} is exponentially 
decreasing function of the space separation $\vec x^2$ which can be parameterized 
at large distances as 
\be 
\Phi_1(\vec x^2) \sim \Phi\, \e^{-\frac{|\vec x|}{\lambda}}\,,\qquad 
\frac{d\Phi_2}{d x^2}(\vec x^2) \sim \Phi'\, \e^{-\frac{|\vec x|}{\lambda}} 
\label{ansatz} 
\ee 
with $\lambda$ being the correlation length,  $\Phi$ and $\Phi'$ being the 
overall normalization scales. This ansatz leads to the following 
expression for the function $\tilde \Phi(\sigma)$ defined in \re{22} 
\be 
\tilde \Phi_1(\sigma)=\frac{\Phi}{\sqrt{4\pi\lambda^2}}\,\sigma^{-3/2} 
\exp\lr{-\frac1{4\sigma\lambda^2}} 
\label{spectr} 
\ee 
and to similar expression for the function $\tilde \Phi_2(\sigma)$. 
The lattice determination of these scales gives \cite{G} 
\be 
\Phi=7.2\times 10^8\, \Lambda_L^4\,,\qquad \lambda=1/(183\, \Lambda_L) 
\label{vals} 
\ee 
with the lattice QCD parameter 
\be 
\Lambda_L= (0.005\pm 0.0015)\ {\rm GeV}. 
\ee 
In addition, $|\Phi'|\approx \Phi/10$ and therefore we may neglect the 
contribution of the $\Phi_2-$function to \re{28}. Then, substituting the ansatz 
\re{ansatz} into the definition \re{30} and \re{30-0} we get the following expressions 
for the coefficients 
\be 
c_1= \frac{\Phi \lambda^2}{2N}=3584\, \Lambda_L^2 \,,\qquad 
c_0 = \frac{2 \Phi \lambda^3}{N}=78 \, \Lambda_L. 
\label{c0c1} 
\ee 
Their values depend strongly on the lattice scale $\Lambda_L$. Choosing this 
scale at the upper boundary $\Lambda_L=0.0065\,{\rm GeV}$, or equivalently 
$\lambda=0.17\  {\rm fm}$, and 
using \re{vals} we estimate the scales \re{c0c1} as 
\be 
c_1^{(\rm latt)} = 0.15\ {\rm GeV}^2\,, \qquad c_0^{(\rm latt)} = 0.51 \ {\rm GeV}. 
\label{cent} 
\ee 
Let us compare our predictions with the phenomenological 
parameterization Eqs.~\re{mod}--\re{5}. To this end we rewrite \re{mod} in equivalent form 
keeping the $\ln Q-$dependent part only 
\be 
\ln W_0^{\rm (phen)}(b^2,Q^2)=-S_{\rm nonPT}= -\ln\frac{Q^2}{Q_0^2}\frac{g_2}2 b^2 . 
\label{phen-W} 
\ee 
We observe that this phenomenological expression matches \re{sol-1} at small $b^2$ 
as 
\be 
c_1^{\rm (phen)} = \frac{g_2}2 = 0.17\ {}^{+0.04}_{-0.04} \ ({\rm GeV})^2. 
\ee 
This expression is in agreement with our calculation \re{cent}. 
 
It follows from our analysis that the parameterization \re{4} 
and \re{phen-W} of nonperturbative corrections to the Sudakov form factor is correct 
only for small $b^2 \ll \lambda^2$, or equivalently for the transverse momenta 
$q^2 \gg \lambda^{-2}$. We would like to stress that although the numerical 
expressions \re{vals} are sensitive to the particular ansatz for the nonlocal 
correlator \re{22}, the general form of the small $b^2$ expansion \re{small} is 
uniquely fixed. According to \re{sol-2}, we expect that at large $b^2$ 
(or small transverse momenta $q^2 \ll \lambda^{-2}$) the asymptotic behaviour 
of the Wilson loop, and as a consequence nonperturbative correction to the Sudakov 
form factor, should have different form \re{sol-2} with the scale $c_0$ 
given by \re{30-0} and \re{vals}. 
 
\begin{figure}[ht] 
\centerline{\epsfig{file=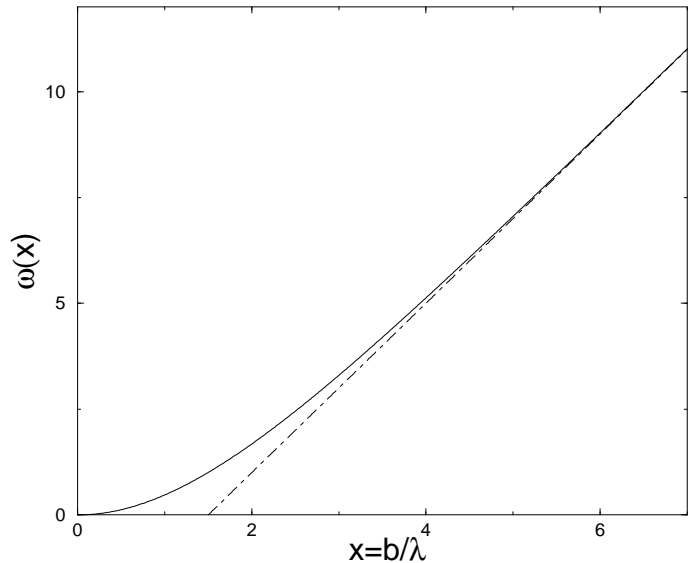,angle=-90,width=9cm}} 
\caption[]{The profile function $w(b/\lambda)$ governing the $b-$dependence of 
 nonperturbative contribution to the Sudakov form factor \re{prof}. The dash-dotted 
 line describes the asymptotic behaviour \re{sol-2}.} 
\label{Fig-2} 
\end{figure} 
 
To summarize, the nonperturbative corrections to small momentum distribution of 
lepton pairs in the Drell-Yan process are described by the expression \re{28} 
which is valid up to $Q-$independent corrections. The asymptotic behavior of 
the exponent of \re{28} at large and small impact vectors is given by \re{sol-1} and 
\re{sol-2}, respectively. Combining together \re{28} with nonperturbative ansatz 
for the nonlocal gauge field correlator, \re{ansatz}, one can find the expression for 
the nonperturbative factor that is valid for an arbitrary $b^2$. 
Another advantage of the ansatz \re{spectr} that integration in \re{28} can be 
performed explicitly leading to the following expression for the nonperturbative 
factor 
\be 
\ln W_0(b^2,Q^2) =- \ln\frac{Q^2}{Q_0^2}\ \frac{\Phi\lambda^4}{N} w(b/\lambda), 
\label{prof} 
\ee 
with the profile function 
\be 
w(x)=\e^{-x}\lr{3+x}-3+2x 
\ee 
which is plotted in Fig.~2. Going over to the 
transverse momentum space we define nonperturbative weight function 
$f_{\rm nonPT}(k^2,Q^2)$ as 
\be 
f_{\rm nonPT}(k^2,Q^2) = \int \frac{d^2 b}{(2\pi)^2}\ 
\e^{i\vec b\cdot \vec k}\ W_0(b^2,Q^2) 
\label{shape} 
\ee 
Then, inserting the factor $W_0(b^2,Q^2)$ into the integrand of \re{1} we find 
that nonperturbative effects produce a smearing of the perturbative $q-$distribution, Eq.~\re{1}, 
with the weight function $f_{\rm nonPT}(k^2,Q^2)$. Substituting 
\re{28} into \re{shape} and taking into account \re{ansatz} we obtain the 
$k-$dependence of the weight function $f_{\rm nonPT}(k^2,Q^2)$ as shown in Fig.~3. 

\begin{figure}[ht] 
\centerline{\epsfig{file=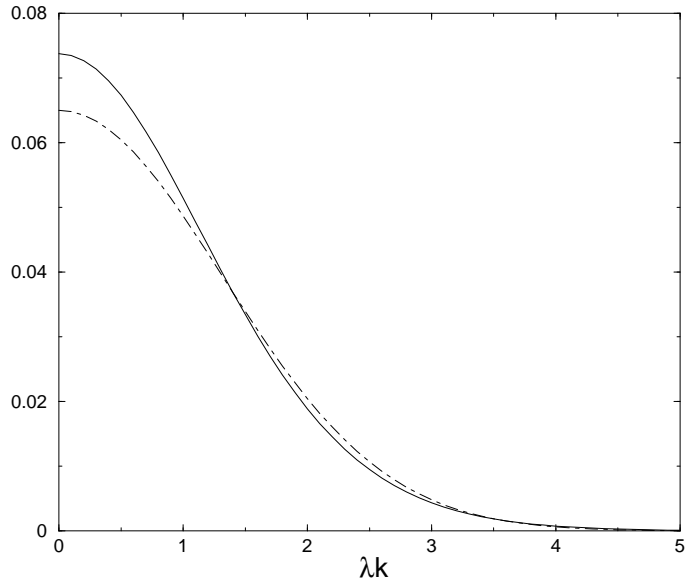,angle=-90,width=9cm}} 
\caption[]{The nonperturbative primordial distribution $f_{\rm nonPT}(k^2,Q^2)$ at 
$Q=91 \ {\rm GeV}$ and $Q_0=1.6 \ {\rm GeV}$. The dash-dotted line describes the 
Gaussian distribution corresponding to \re{sol-1}.} 
\label{Fig-3} 
\end{figure} 
 
\section{Conclusion} 
 
In this paper we have studied nonperturbative corrections to the 
transverse momentum distribution of vector bosons in the Drell-Yan 
process. Factorizing out the Sudakov effects due to soft gluons we 
were able to express their contribution to the distribution in the 
form of the vacuum averaged Wilson loop operator. We have 
demonstrated that the Wilson loop take into account both 
perturbative and nonperturbative corrections. The former are 
resummed through the evolution equation whereas the later define 
the boundary conditions for their solutions. Analyzing large order 
perturbative corrections to the Sudakov form factor we found that 
perturbation theory provides an ambiguous contribution to power 
corrections $\sim(b^2\Lambda^2)^n$ associated with the IR 
renormalons. The IR renormalon contribution suggests the general 
structure of nonperturbative corrections and it can be absorbed 
into nonperturbative boundary conditions for the Wilson loop. 
 
We calculated the nonperturbative contribution to the Sudakov form factor using 
the expansion of the Wilson loop over vacuum fields supplemented with the 
expression for nonlocal gauge invariant field strength correlator. Although 
the Wilson loop is defined in an essentially Minkowski kinematics, the 
part of the nonperturbative contribution depending on the invariant mass of the 
produced vector boson $Q^2$ is governed by asymptotics of the correlator at large 
space-like (Euclidean) separations and therefore can be calculated using 
conventional nonperturbative methods. Applying the results of lattice 
calculations we found that the obtained expression for the nonperturbative power 
corrections is in qualitative agreement with known phenomenological 
expressions at large transverse momenta and deviate from them at small transverse 
momenta. 
 
\section*{Acknowledgements} 
 
I am most grateful to G.~Korchemsky for numerous stimulating 
discussions and suggestions. I would also like to thank T.~Binoth, 
J.-Ph.~Guillet and E.~Pilon for useful conversations and 
G.~Sterman for critical reading of the manuscript. 
 
This work was supported in part by the EU network 
`Training and Mobility of Researchers', contract  FMRX--CT98--0194.

\end{document}